\documentclass[aps,prl,reprint,groupedaddress,showpacs]{revtex4-1}

\usepackage{graphicx}
\usepackage{epstopdf}
\usepackage{wasysym}
\usepackage{color}
\usepackage{ulem} 

\newcommand{\ket}[1]{\ensuremath{\mid\!\! #1 \rangle}}
\begin{document}

\title{Electrical read out for coherent phenomena involving Rydberg atoms in thermal vapor cells}

\author{D. Barredo}
\affiliation{5. Physikalisches Institut, Universit\"{a}t Stuttgart, Pfaffenwaldring 57, 70550 Stuttgart, Germany}
\author{H. K\"ubler}
\affiliation{5. Physikalisches Institut, Universit\"{a}t Stuttgart, Pfaffenwaldring 57, 70550 Stuttgart, Germany}
\author{R. Daschner}
\affiliation{5. Physikalisches Institut, Universit\"{a}t Stuttgart, Pfaffenwaldring 57, 70550 Stuttgart, Germany}
\author{R. L\"ow}
\affiliation{5. Physikalisches Institut, Universit\"{a}t Stuttgart, Pfaffenwaldring 57, 70550 Stuttgart, Germany}
\author{T. Pfau}
\affiliation{5. Physikalisches Institut, Universit\"{a}t Stuttgart, Pfaffenwaldring 57, 70550 Stuttgart, Germany}

\date{\today}

\begin{abstract}
{We present a very sensitive and scalable method to measure the population of highly excited Rydberg states in a thermal vapor cell of rubidium atoms. We detect the Rydberg ionization current in a 5 mm electrically contacted cell. The measured current is found to be in excellent agreement with a theory for the Rydberg population based on a master equation for the three level problem including an ionization channel and the full Doppler distributions at the corresponding temperatures. The signal-to-noise ratio of the current detection is substantially better than purely optical techniques.
 }
\end{abstract}


\maketitle

Coherent phenomena involving strongly interacting Rydberg atoms have recently led to the demonstration of first quantum devices like quantum logic gates \cite{Grangier,Saffman,Saffmancnot} and single photon sources \cite{Kuzmich} based on ultracold atoms. All these experiments require precise control over the highly excited states populations, which can be probed directly by field ionization \cite{Gallagher,Pillet} or by fluorescence techniques involving Rydberg shielding \cite{Bloch}. Since the strong vdW interaction has recently also been observed in vapor cells \cite{Baluktsian}, scalable quantum devices based on the Rydberg blockade in above room temperature ensembles seem to be also within reach \cite{newsandviewsRLTP}. However, ion detectors as electron multipliers or multi-channel plates cannot be used in dense thermal vapors. For this reason, in thermal cells, most studies today use an indirect measurement of the excited state population by analyzing light fields leaving the atomic ensemble. Nevertheless, it is desirable to study not only the back-action of the vapor on the light, typically via electromagnetically induced transparency (EIT) \cite{Mohapatra}, but also to measure directly the number of excited Rydberg states. One method, developed almost a century ago \cite{Kingdon,Hertz}, makes use of thermionic diodes \cite{Stoicheff,Niemax,Weis}. There, one of the electrodes is heated to emit electrons, which produce space charge limited gain for the amplification of ionized Rydberg atoms. The need of long ion trapping times requires large geometries for the space charge region, and an additional shielded excitation region to minimize the effect of disturbing electric fields during excitation of the highly polarizable Rydberg atoms. Despite its high sensitivity, this drawback sets a practical limitation for further applications where size and scalability play a role.

Here we demonstrate that, in a symmetric configuration of atomic vapor between two transparent field plates, sizable currents in the nA regime reflect directly the Rydberg population and can be used as a probe with very good signal-to-noise ratio. This opens unique possibilities to probe very efficiently small spectroscopic features involving Rydberg states in thermal vapor but also might be used to stabilize lasers. By extending this concept to an array of pixel-wise arranged electrodes, high resolution spatial information on the Rydberg population can be obtained.

\begin{figure}
\includegraphics[width=\linewidth]{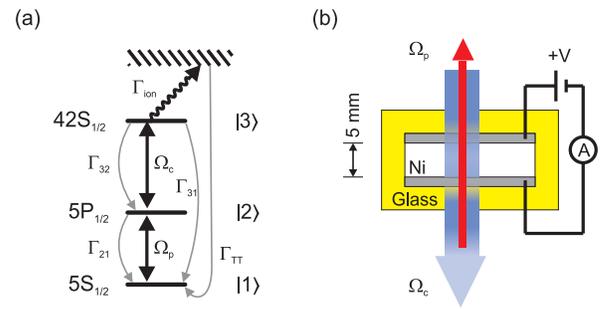}
\caption{(a) Three-level excitation scheme. The probe beam of Rabi frequency $\Omega_{p}$ couples the states $\ket{1}$ and $\ket{2}$, while the control laser of Rabi frequency $\Omega_{c}$ couples the states $\ket{2}$ and $\ket{3}$. (b) Experimental set-up. The excitation takes place inside a glass cell filled with Rb vapor. Rydberg ions are detected as a current between two electrodes held at a slightly different electric potential $V$. The electrodes consist of 5 $nm$ thin metallic layers of Ni which provide a 75\% transparency for the excitation lasers.
\label{fig1}}
\end{figure}

The experiments were performed with the setup schematically shown in Fig. \ref{fig1}. The Rb vapor is confined in a $l =$ 5 mm electrically contacted cell which has been described in detail elsewhere \cite{Daschner}. In essence, it consists of two glass substrates glued to the edges of a glass frame attached to a reservoir tube. Two thin (ca. 5 nm) layers of Ni sputtered on the substrates act as field plates and provide electrical contact to the outside while still allowing for a high (ca. 75\%) optical transparency for the excitation lasers. Inside the cell, ${}^{85}\text{Rb}$ atoms are excited to a Rydberg state in a two-photon process in an EIT scheme \cite{Mohapatra}. A light field at $\lambda_{p} \sim 795$ nm of Rabi frequency $\Omega_{p}$ couples the $5S_{1/2} F = 3$ ground state and the $5P_{1/2} F' = 2$ intermediate state. In the second excitation step a laser at $\lambda_{c} \sim 474$ nm populates the $42S$ Rydberg level. Both the probe and the coupling lasers cross the cell perpendicular to the field plates and overlap in the excitation region in a counter-propagating manner with beam waists ($1/e^2$ radii) $w_p=0.16$ mm and $w_c=0.51$ mm, respectively. Highly excited atoms are ionized mainly by collisions with other atoms and the ionization products are detected as a current flow between the two electrodes, in which a low bias voltage (usually $\sim 0.1$ V) is applied. This low voltage does not perturb significantly the highly polarizable Rydberg atoms. The cell is enclosed in an electrostatically shielded oven and both the temperature of the cell and the reservoir can be varied independently. The cell was always kept at a higher temperature with respect to the reservoir to avoid condensation of Rb on the glass surface and the atom density is controlled by regulating the reservoir temperature. Finally, the transmission of the probe beam and the ionization current are measured simultaneously with a photodiode (Thorlabs DET36A) and a current amplifier (Keithley 427), respectively.

\begin{figure}
\includegraphics[width=\columnwidth]{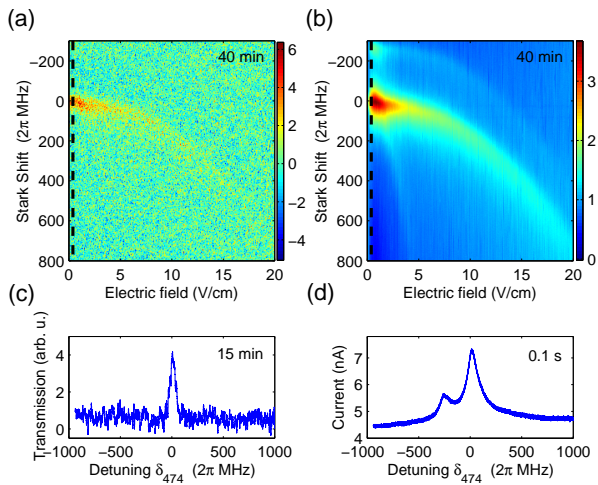}
\caption{Stark maps of the 42S Rydberg state. Transmission of the probe field (a) and ionization signal (b), measured simultaneously, after background subtraction.(c) Averaged optical signal measured with a lock-in amplifier and the corresponding single-shot ionization signal (d) recorded 9000 times faster.
\label{fig2}}
\end{figure}

One of these measurements is shown in Fig. \ref{fig2}. It corresponds to a Stark map for the 42S Rydberg state, where the coupling laser is scanned around the two-photon resonance. The intensity of the coupling laser is modulated with an AOM and the optical signal (Fig. \ref{fig2} (a)) is measured with a lock-in amplifier. The strength of the electric field is controlled by ramping the voltage applied directly to the electrodes. The corresponding ionization current signal is shown in Fig. \ref{fig2} (b), after background subtraction. This background is mainly due to photoelectrons produced by the 474 nm light on the Ni electrodes, whose work functions are lowered by adsorbed Rb atoms on the surfaces. In the optical signal the EIT peak is clearly observed for low voltages but broadens and is hardly observable for $V > 10$ V/cm. In contrast, the ionization current produces a map with a higher contrast in which the Rydberg signal can be followed over the full range of the measurement. The second feature starting at a Stark shift of $\sim -250 \times 2\pi$ MHz corresponds to Rydberg excitation from the Doppler-allowed $5P_{1/2} F' = 3$ hyperfine state \footnotemark{}, which is barely visible in the EIT signal.
  Such a low optical signal may be surprising if compared to that obtained with longer vapor cells and therefore, it is worth stressing that similar signal-to-noise ratios have been reported for this type of cells \cite{Daschner}, where peak heights are further reduced by preasure broadening ($\sim 50 \times 2\pi$ MHz). The striking difference in the signal-to-noise ratio between the optical and the ionization signal is better seen in a cut at a constant voltage. In Fig. \ref{fig2} (c) the measured EIT peak at a voltage of 0.4 V/cm was averaged over 90 traces at a scanning rate of 100 mHz, with an integration time constant of 3 ms. In comparison, the ionization signal (d) was obtained in a single-shot with the coupling laser scanning at 10 Hz and a rise time constant of the current amplifier of 10 $\mu$s. Measuring ions is experimentally more favorable because ionization signals can in principle be extracted from an essentially zero background, in contrast to optical detection. In practice, in this excitation scheme, electrons produced by photoelectric effect and two-photon Rb ionization signals contribute to the observed background and set the ultimate limit to which noise can be reduced \cite{Weis}.

\begin{figure}
\includegraphics[width=\columnwidth]{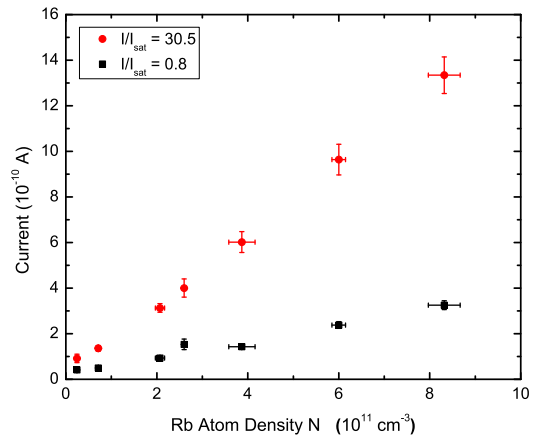}
\caption{(a) Scaling of the measured ionization signal with ground state density for two different probe laser intensities $I_{sat} = 0.8$ and 30.5. Error bars represent $\pm 1$ standard deviation  confident intervals.
\label{fig3}}
\end{figure}

To demonstrate that this electrical current is a direct measure of the population of the 42S state, ionization spectra were first measured as a function of Rb vapor density for various probe laser intensities (Fig. \ref{fig3}). For each point the atom density $N$ was estimated from control absorption measurements and peak heights were extracted from fittings to lorentzian profiles. Fig. \ref{fig3} shows a linear scaling for $N >2 \times 10^{11} cm^{-3}$, indicating that the probability that a Rydberg atom gets ionized in a collision stays constant. This is the expected behavior if the collisional ionization rate $\Gamma_{ion}$ exceeds the sum of the rates of all non-ionizing decay channels. For high Rydberg states, with long radiative lifetimes this condition can be fulfilled even for low background gas pressures \cite{Niemax}. We have ruled out surface ionization as the origin of the detected current. Surface ionization should produce a velocity selection and subsequently an asymmetry in the Doppler profile distribution as a consequence of the geometrical arrangement. Such asymmetry was not observed. Photoionization due to the 474 nm light was also discarded because the signal was not observed to increase if a third strong non-resonant 480 nm laser is added. Also, ionization rates by blackbody radiation are expected to be very small \cite{Beterov}. Collision with electrons should also play a minor role in ionization, as judged from experimental ionization cross sections \cite{Kashtanov}, which lead to a upper limit for ionization in our conditions of about 5 kHz. This is further supported by the fact that an increase of the non-resonant light intensity increases the current background but does not produce higher signals.

The next step was to investigate the behaviour of the signal as a function of laser power. To do so we performed a simulation based on the analytical solution of Lindblad equation in steady state, in which all coupling strengths and detunings are included as fixed parameters.
 \begin{equation}
\dot{\rho}=-\frac{i}{\hbar}\left[\hat{H},\hat{\rho}\right]+\hat{L}(\rho)=0
\end{equation}
with the Hamilton operator
    \begin{eqnarray*}
        \hat{H}=\hbar\left(\begin{array}{ccccc} 0 & \frac{1}{2}\Omega_{p} & 0 & 0 \\
            \frac{1}{2}\Omega_{p}^\ast & -\delta_{p} & \frac{1}{2}\Omega_{c} & 0\\
            0 & \frac{1}{2}\Omega_{c}^\ast& -\delta_{p}-\delta_{c} & 0 \\
            0 & 0 & 0 & 0
                \end{array}\right)
\end{eqnarray*}
where $\Omega_{p}$, $\Omega_{c}$,  $\delta_{p}$ and $\delta_{c}$ are the probe and coupling Rabi frequencies and energy detunings, respectively.  The velocity distribution of the atoms is taken into account by their Doppler shift $\delta_{\text{v},i}=\vec{k_{i}}\times v$ for each transition, weighted by a 1D Boltzmann distribution $g_v=\sqrt{m/(2\pi k_B T)}\times\text{exp}({-m v^2/(2 k_B T)})$.
The Lindblad operator $\hat{L}(\rho)$ includes the decays $\Gamma_{ij}$ as depicted in Fig. \ref{fig1}, where the ionization channel is essentially treated as a forth level with $\Gamma_{34} = \Gamma_{ion}$, and the cycle is closed with the transit-time rate $\Gamma_{TT} = 0.25 \times 2\pi$ MHz, estimated from the beam waist $w_p$ and the average atom velocity at a mean cell temperature of $\sim 70^\circ C$ \cite{Sagle}. The ionization rate was estimated from the observed width of the EIT signal to be $\Gamma_{ion} = 50 \times 2\pi$ MHz. This is consistent with measurements performed in similar cells \cite{Daschner} and comes presumably from pressure broadening due to contaminants inherent to these glued-type vapor cells.

\begin{figure}
\includegraphics[width=\columnwidth]{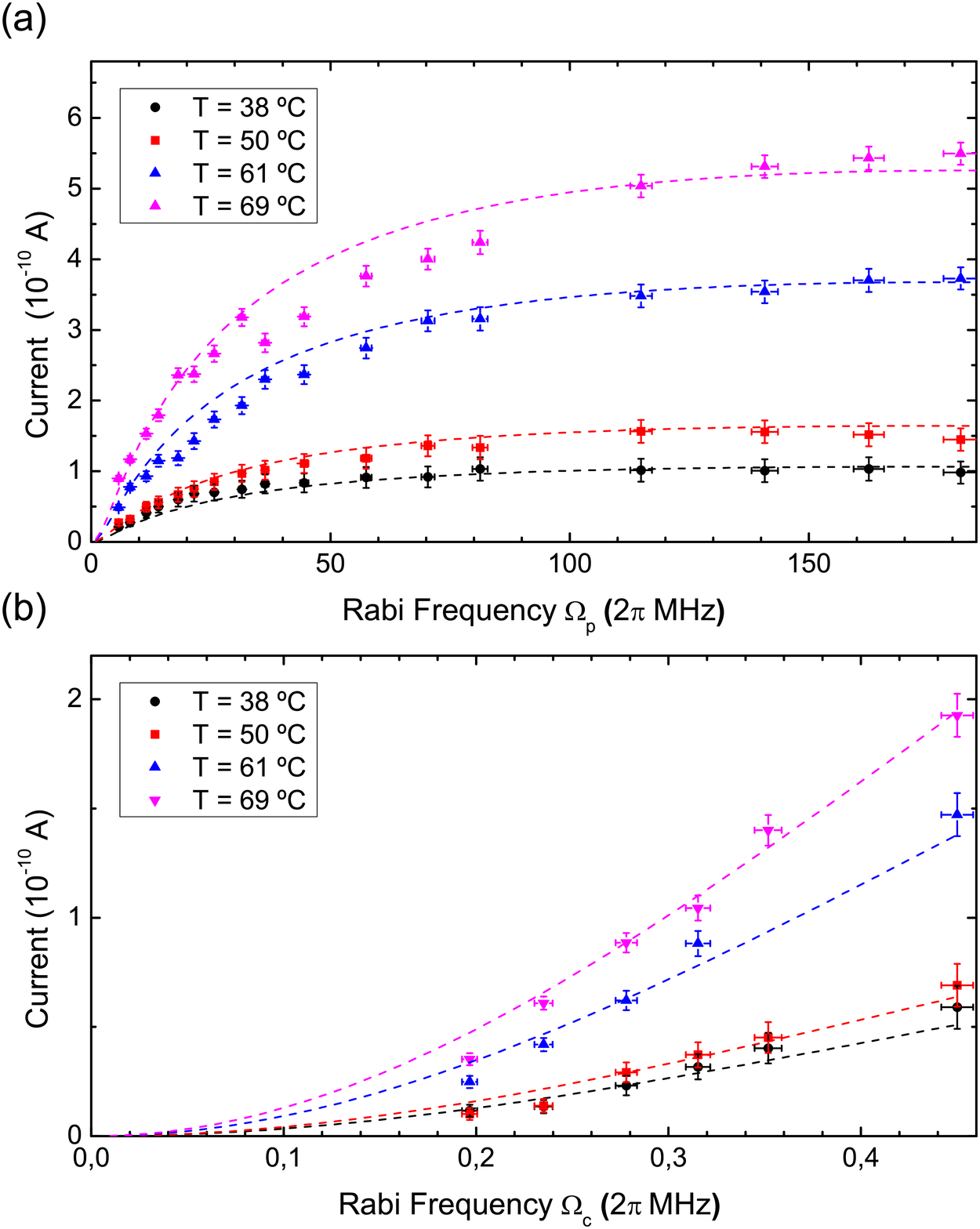}
\caption{Measured ionization signal as a function of the peak probe field Rabi frequency $\Omega_{p}$ (a) and coupling laser peak Rabi frequency $\Omega_{c}$ (b) for different ground state atom densities. Dashed lines are theoretically calculated Rydberg state populations scaled to the corresponding atomic density. Error bars represent $\pm 1$ standard deviation confident intervals.
\label{fig4}}
\end{figure}

The comparison between the theoretical calculation and the experimental results is summarized in Fig.~\ref{fig4}. The scaling behavior of the current with both the probe (a) and coupling laser (b) Rabi frequencies is shown. Each dashed line represents the calculated population scaled with a prefactor. The Rabi frequencies were estimated from the beam waists at the excitation region and the incident laser power. No further correction was made due to the attenuation of the radiation in the vapor. For this reason only the low density curves are shown. In these conditions, and since most Rabi frequency values in Fig.~\ref{fig4} (a) correspond to probe intensities much higher than the saturation intensity of the $5S_{1/2} - 5P_{1/2}$ transition, propagation effects are expected to be small. For higher densities propagation of the light field in the optically thick medium must be included in the model \cite{Vanier}, and the experimental data depart from the prediction of our simple model. A good agreement is found in both datasets, demonstrating that the measured ionization signal can be directly correlated with the Rydberg state population. A rough check for the order of magnitude of the prefactors used to scale the Rydberg population to the measured current can be obtained from our simulation. The highest current value in Fig. \ref{fig4} (a) of $\sim$ 5 $\times 10^{-10}$ A ($\Omega_p \sim 180 \times 2\pi$ MHz, $\Omega_c \sim 0.45 \times 2\pi$ MHz) corresponds to a steady state population of the Rydberg level $\rho_{33} \sim 4 \times 10^{-6}$, including Doppler averaging over the velocity classes. The expected current from the experimental parameters, i.e., the ionization volume ($l\times \pi w_p^2$), the $5S_{1/2} F = 3$ ground state atom density ($N_g = 9.4 \times 10^{10} cm^{-3}$), and assuming $\Gamma_{ion} = 50 \times  2\pi$ MHz as the Rydberg production rate, yields a value of about 8 nA, which is about a factor ten higher than what we measure. Therefore, a gain mechanism due to space charge limitation or avalanche ionization in the medium is not evident.

In conclusion, we have shown that the measurement of the Rydberg ionization current in thermal vapor cells provides, over a wide range of densities and laser powers, a direct measure of Rydberg states population. Since the Rydberg blockade fraction and the probe beam optical susceptibility are coupled by a simple universal relation \cite{Pohl}, the combination of EIT spectroscopy with the measurement of the ionization current represents a promising approach to study interaction effects between Rydberg atoms in a thermal environment \cite{AdamsUniversal}. This method is not only well suited for applications in quantum information processing based on the blockade of the Rydberg population, but also enables applications in precision spectroscopy and laser stabilization. Moreover, spatial resolution and scalability are possible, as the electrodes can be structured similar to the way it is done in display technology, and Rydberg atoms can be excited in much smaller vapor cells \cite{Kuebler}.

\begin{acknowledgments}
We acknowledge helpful discussions with J. P. Shaffer. We thank E. Kurz and N. Fr\"uhauf for assistance in the construction of the cell and E. Sch\"oll for additional data integrity checks. This work was supported by a Marie Curie Intra European Fellowship within the 7th European Community Framework Programme (272831), by the ERC under contract
number 267100 and BMBF within QuOReP (Project 01BQ1013). We also acknowledge the financial support of the Future and Emerging Technologies (FET) programme within the Seventh Framework Programme for Research of the European Commission, under FET-Open grant MALICIA (265522).
\end{acknowledgments}

\bibliographystyle{prsty}

\footnotetext[1]{Note that, due to different Doppler velocity class selection by the probe and coupling lasers, the hyperfine splitting of the $5P_{1/2}$ is scaled by the factor $(\lambda_{p}/\lambda_{c} - 1)$.}

\end{document}